# Deployment of physics simulation apps using Easy JavaScript Simulations

Félix J. García Clemente
Dpto. Ingeniería y Tecnología de Computadores
Universidad de Murcia, Spain

Francisco Esquembre
Dpto. Matemáticas
Universidad de Murcia, Spain

Loo Kang Wee
Education Technology Division
Ministry of Education, Singapore

*Abstract*— **Physics simulations are widely used to improve the learning process in science and engineering education. Deployment of a computational physics simulation/model is extremely complex given the fact that both knowledge and skills for the science equations and the computational and programming aspects are required for a fully functional simulation, typically requiring a science educator and computer scientists/developer to work together. However, when using Easy JavaScript Simulation (EjsS) modeling toolkit, the instructor can be both the science educator and computer programmer, only needing to define the simulation variables, model and view, and the modeling toolkit can generate the computer codes for the physics simulation. Moreover, the programming aspects can become even more complex if simulations require being optimized for both Android and iOS mobile devices. The current version of EjsS provides instructors with an authoring tool that includes facilities for the creation of such JavaScript simulations as mobile apps, thus simplifying the purely programming aspects. This paper presents a new and novel EjsS functionality to generate physics simulation apps for iOS and Android. The generation process is based on the integration of the Ionic/Cordova and AngularJS technologies into EjsS. Finally, we present several working examples based on the works of Open Source Physics at Singapore.**

## I. INTRODUCTION

Nowadays, physics simulations are widely used to improve the learning process in science and engineering education. There are many advantages to using simulations to teach science and engineering processes. Firstly, learners can observe science processes and interact with input parameters to explore and understand the model or system's behaviour. Secondly, learners can see multiple representations and visualize the models that may not actually be seen in real time and space [1]. And thirdly, simulations can serve as virtual laboratories where learners can manipulate science and engineering processes, that maybe happening very fast thus, difficult to observe in real-life context, with lower costs and no risks.

While the advantages of using simulation is well documented in research and practice, the deployment of such a physics simulation is not trivial. In the creation of a simulation, the instructor will need to define the simulation variables, model and view, and he/she must also focus on the programming aspects, usually too complex for science educators. The alternative most instructors choose to do is to use third-party simulations that can be downloaded from free online repositories (e.g. OSP Collection [2] and PhET Interactive Simulations [3]). We argue that for deeper integration of learning experiences necessary for preparing next generation learners, the educator will want to adapt these simulations to better meet the specific learning outcomes. In addition to adapting simulations, the instructors will need these same simulations optimized for mobile devices in order to take advantage of their features and provide learners with a better experience.

Easy JavaScript Simulations (EjsS) provides instructors with a click and drag programmer interface on an authoring tool that includes facilities for the creation of JavaScript simulations, generate the computer codes automatically, thus simplifying the purely programming aspects. The instructor defines the physics model and links model definitions to the components of the view. EjsS includes the automatic generation of code that integrates the simulation definition into the EjsS JavaScript framework. The generated simulation can then be used in any web browser with HTML5 and WebGL support (currently, most of them) in any kind of device, such as smartphones or tablets.

This paper presents new EjsS functionality is capable of generating educational physics simulation apps for Android and iOS. This generation process is based on the integration of the Ionic/Cordova and AngularJS technologies into EjsS. Finally, we present several working examples based on the works of Open Source Physics at Singapore.

## II. DEPLOYMENT OF PHYSICS SIMULATION APPS

### A. EjsS Authoring Tool

EjsS Authoring Tool is addressed to non-expert programmers, although authors need a basic knowledge of JavaScript programming [4]. The author keys in the physics equations in the model, designs the graphical User Interface (UI) using drag and drop and links model functions to UI events (see Figure 1).

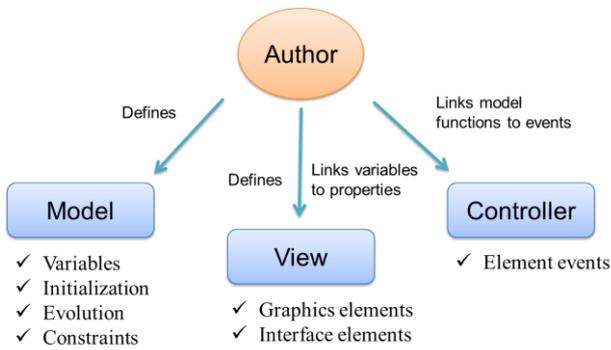

Fig. 1. EjsS Model-View-Controller (MVC) simplified schematics of how the authoring toolkit works.

EjsS Authoring tool generates simulations following a Model-View-Controller (MVC) pattern that is internally implemented in EjsS Library [5]. EjsS Library defines a simulation flow that allows the evolution of any physical process. Figure 2 shows the simulation flow diagram and where the author can define each step into the EjsS Authoring Tool (i.e. Variables, Initialization, Fixed relations, and Evolution with Event Loop and Timer Loop for the time step of the simulation).

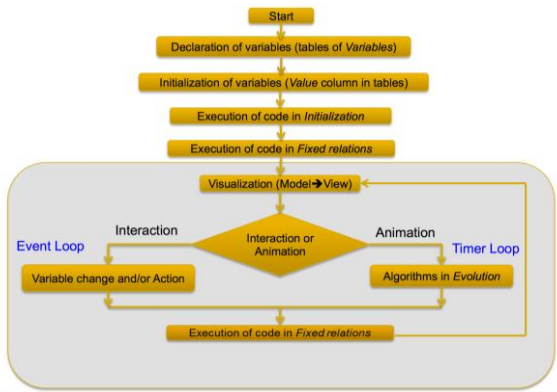

Fig. 2. EjsS simulation computational flow diagram.

EjsS Authoring tool creates either a ZIP archive of XHTML pages with JavaScript or an ePub. The simulation author can upload these simulations to his/her web server and open them with his/her web browser or ePub reader.

### B. Generation of simulation apps

There are two alternatives to develop simulation apps: native and hybrid apps. Native apps are designed and coded for a specific kind of device, however hybrid apps are built using cross-platform web technologies, such as HTML5, CSS and Javascript. In our case, we want to create apps for Android and iOS, and still maintain the EjsS MVC pattern and the simulation flow as well as EjsS Library. Therefore, we opted to generate hybrid apps based on a simple template that can be easily modified using any HTML editor.

In the hybrid apps development world, there are several different appropriate technologies, among them, Cordova, PhoneGap and Ionic [6]. Ionic is built on top of AngularJS and Apache Cordova, and provides tools and services for developing hybrid mobile apps using web technologies like CSS and HTML5. In our opinion, Ionic presents the best features, therefore the EjsS generation of simulation apps uses Ionic platform of which the EjsS modeling tool generates the Ionic template for serving the simulation on the app.

The generation process consists of a few steps that end with the further building of the app in the platforms Android and iOS. The first step is to install the platform Ionic in your computer [7]. This step needs to be done only once, no matter how many Ionic apps are created.

The second step is to create an Ionic project using the Ionic tool. The tasks below need to be done once per simulation app.

- Create a new blank Ionic project:

```
> ionic start MyModelApp blank
```

- Add target platforms iOS and Android:

```
> ionic platform add ios
> ionic platform add android
```

What these commands do is to create an Android and iOS folder and these contain the Android and XCode projects for the simulation app. These projects will eventually be built into the packages that will be deployed to the respective App Stores.

The third step is to generate a prepackaged model, i.e. a ZIP file that is obtained with the "Prepackage for App" option of EjsS menu (see Figure 3). And then, we will extract the prepackaged files into the *www* folder of our black Ionic project.

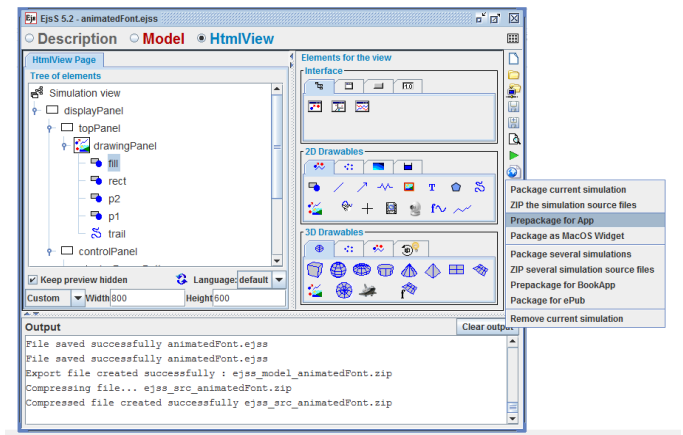

Fig. 3. Generation of a prepackage for a simulation app in EjsS.

Finally, we could test and build the app using Ionic tools. In addition, we may want to edit the Ionic files of the project in order to change the name, description and author entries, or change the icon and splash images.

Thus, we have briefly described how the EjsS toolkit enables the generation process of an app ZIP file that can be used to deploy on Android and iOS as mobile apps, which we argue is so much more accessible to educators now requiring no prior knowledge about mobile apps programing/development.

## C. Structure of EjsS apps

The structure of EjsS simulation apps is based on the Ionic side menu template. Once the project is created, you can find the *index.html* in the root of *www* folder. This file contains our side menu code. The following code shows the main tags for our side menu.

```
<ion-side-menu side="left">
  <ion-header-bar class="bar-stable">
    <h1 class="title">{{app_menu_title}}</h1>
  </ion-header-bar>
  <ion-content scroll="true">
    <ion-list>
      <div ng-repeat="entry in pages">
        <ion-item … >{{entry.title}}</ion-item>
      </div>
    </ion-list>
  </ion-content>
</ion-side-menu>
```

The *ion-header-bar* adds a fixed header bar above the side menu and the *ion-content* includes the *ion-list* with the menu items. Note the *app_menu_title* and *entry.title* variables must be linked to string values.

In addition, the *index.html* includes the description of the top bar with navigation buttons and an iframe with the content. The following code shows the part related to the content.

```
<script id="model_page.html" type="text/ng-template">
  <ion-view>
    <ion-content scroll="true" overflow-scroll="true" class="iframe-wrapper">
      <iframe data-tap-disabled="true" src={{currentPage.url}}>
      </iframe>
    </ion-content>
  </ion-view>
</script>
```

The iframe is used to display the web page associated to the menu item selected and it is referenced by the *currentPage.url* variable. Through AngularJS the web pages are linked to the side menu. The following code shows the content of the file *pages.js* that includes the variables used by AngularJS.

```
var app_title = "MyModelApp";
var app_menu_title = "Contents";
var app_toc = [
 {
  title: "Simulation", type: "model_page",
  url: "model_pages/myModelApp_Simulation.xhtml"
 } ,
 {
  title: "References", type: "references",
  url: "other_pages/references.html"
 }
];
```

Our template has two additional directories to store the content. The *model_pages* directory includes the simulation and the *other_pages* does the information related to the copyright and additional information.

Any author with basic knowledge about web programing could modify the template and adapt it to his/her needs. Even an advanced author could integrate the simulation into other template (e.g. Ionic tabs template) or create a new Ionic template.

## III. EjsS Apps by Open Source Physics at Singapore

In this section we present several EjsS simulation apps developed by Open Source Physics at Singapore (OSP@SG). OSP@SG is a digital library containing Java, JavaScript and Tracker resources. The open source codes are freely shared online to support the teaching and learning of Physics. OSP@SG started in 2012 and involvement 100 teachers and approximately 9,800 students, creating an Open Source Physics community.

OSP@SG has customized or co-developed hundreds of EjsS simulation with our global community members and more recently created 30 Android apps and 6 iOS Apps ideal for showcasing this paper's research and designing apps for experiential learning [8]. These apps address from basic concepts like measurements and instruments, while others related to kinematics, gravity and electromagnetism. The following table shows the number of downloads for the latest published apps.

TABLE I. Downloads for several OSP@SG Apps on 14/11/2016

|   | Name of App | Downloads | Date Published |
|---|---|---|---|
| 1 | Micrometer Simulator | 3,338 | 11/08/2016 |
| 2 | Vernier Calipers Simulator | 2,948 | 10/08/2016 |
| 3 | DC Motor 3D Simulator Lab | 717 | 18/08/2016 |
| 4 | AC Generator 3D Virtual Lab | 401 | 18/08/2016 |
| 5 | Oscilloscope Simulator | 205 | 14/08/2016 |

The most downloaded app is *Micrometer Simulator* [9]. This app has most functional physical parts of a real micrometer (see Figure 4). It allows drag action to change the size and position of the object (black) into the anvil and spindle (jaws). The learner can control the position of the spindle and check his/her measurement.

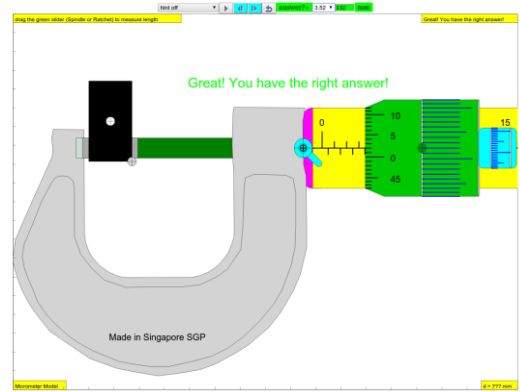

Fig. 4. Micrometer Simulator available on Android and iOS app stores respectively

*DC Motor 3D Simulator Lab* is a simulation app to allow exploration of direct current electrical motors turn electricity into motion by exploiting electromagnetic induction (see Figure 5). A current-carrying loop that is placed in a magnetic field experiences a turning effect. A simple direct current (DC) motor is illustrated here.

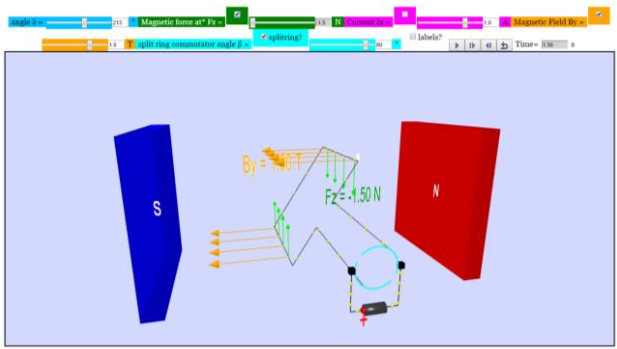

Fig. 5. DC Motor 3D Simulator Lab available on Android and iOS app stores respectively.

*Mass Gravity Virtual Lab* allows the inquiry of gravity concepts of field and potential for a two mass setup (see Figure 6). Every object sets up a gravitational field around itself due to its mass. When two objects enter each other's gravitational fields, they will be attracted towards each other.

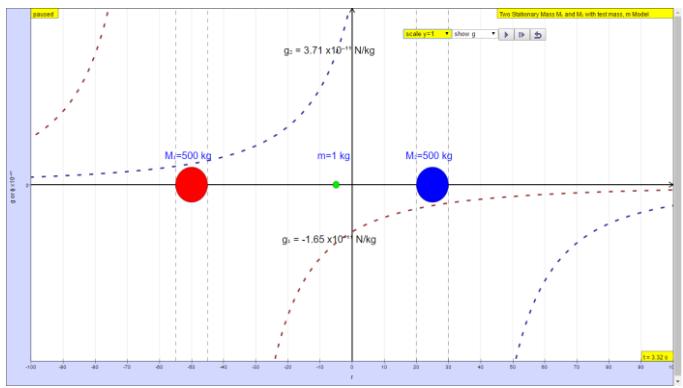

Fig. 6. Mass Gravity Virtual Lab currently available on Android app store.

More apps can be found at the Google Play Store and iOS AppStore by searching the author identification "OpenSourcePhysicsSG" and "Ezzy Chan" respectively.

IV. CONCLUSIONS

This paper argues for deploying simulations for science and engineering education for the advantage of mobile devices features to further enhance experiential learning, possible through using the EjsS modeling tool to generate and eventually deploy as hybrid apps in a few relatively accessible steps using Easy JavaScript Simulations. We discussed core deployment stages such as the EjsS Model-View-Controller (MVC) simplified schematics of how the authoring toolkit works, generation of hybrid mobile apps and the structure of our Ionic template. Finally, we demonstrate several working Google and Apple Apps from our Open Source Physics at Singapore digital library.

REFERENCES

[1] D. Wong, S.P. Poo, N.E. Hock, and W.L. Kang, "Learning with multiple representations: an example of a revision lesson in mechanics," Physics Education, 46 (2), pp. 178, 2011.
[2] Open Source Physics, "OSP Collection," http:// http://www.opensourcephysics.org/.
[3] University of Colorado, "PhET Interactive Simulations," https://phet.colorado.edu/.
[4] University of Murcia, "Easy JavaScript Simulation," http://www.um.es/fem/EjsWiki/.
[5] F. J. Garcia Clemente and F. Esquembre, "Ejss: A javascript library and authoring tool which makes computational-physics education simpler," XXVI IUPAP Conference on Computational Physics (CCP), Boston, USA, 2014.
[6] Ionic, "Ionic Framework," https://ionicframework.com/.
[7] Ionic, "Get started with Ionic Framework," https://ionicframework.com/getting-started/.
[8] L.K. Wee, "One-dimensional collision carts computer model and its design ideas for productive experiential learning," Physics Education, 47, (3), pp. 301, 2012.
[9] L.K. Wee and H.T. Ning, "Vernier caliper and micrometer computer models using Easy Java Simulation and its pedagogical design feature-ideas to augment learning with real instruments," Physics Education arXiv preprint arXiv:1408.3803, 2014